	\documentclass[conference]{IEEEtran}

\usepackage{cite}
\ifCLASSINFOpdf

\else

\fi
\usepackage{subfig}
\usepackage{bm}
\usepackage{amsmath}
\usepackage{epsf}
\usepackage{graphics}
\usepackage{ amssymb }
\usepackage[dvips]{graphicx}
\usepackage{epsfig}
\usepackage{cite}
\usepackage[linesnumbered,ruled,vlined]{algorithm2e}
\usepackage{colortbl}
\usepackage{color}
\usepackage{multicol}
\usepackage{bm}
\usepackage{amsmath}
\usepackage{epsf}
\usepackage{graphics}
\usepackage{ amssymb }
\usepackage[dvips]{graphicx}
\usepackage{epsfig}
\usepackage{cite}
\usepackage[linesnumbered,ruled,vlined]{algorithm2e}
\usepackage{graphicx}
\usepackage{epsfig}
\usepackage{latexsym}
\usepackage{amsfonts}
\usepackage{here}
\usepackage{rawfonts}
\usepackage[utf8]{inputenc}
\usepackage[english]{babel}
\usepackage{amsmath}
\usepackage{amsfonts}
\usepackage{amssymb}
\usepackage{color} %May be necessary if you want to color links
\usepackage{bm}
\usepackage{listings}
\usepackage{caption}
\usepackage{amssymb}
\usepackage{amsthm}
\usepackage{graphicx}
\usepackage{epstopdf}
\usepackage{listings}
\usepackage{float}
\usepackage{amsmath}
\usepackage{amssymb}
\usepackage{amsfonts}
\usepackage{epstopdf}

\usepackage{multirow}
\usepackage{amscd}
\usepackage{mathrsfs}
\usepackage{graphicx}
\usepackage{color}
\usepackage{url}
\usepackage{bm}
\usepackage{setspace}

\usepackage{footnote}
\newtheorem{lemma}{Lemma}
\newtheorem{definition}{Definition}

\usepackage[]{algorithm2e}
\renewcommand{\paragraph}{\@startsection{paragraph}{4}{\z@}%
	{-2.00ex\@plus -1ex \@minus -.2ex}%
	{0.5ex \@plus .2ex}%
	{\normalfont\normalsize}}
	
 \usepackage[protrusion=true,expansion=true]{microtype}

\IEEEoverridecommandlockouts
\usepackage{cite}
\usepackage{amsmath,amssymb,amsfonts}
\usepackage{algorithmic}
\usepackage{graphicx}
\usepackage{textcomp}
\usepackage{xcolor}
\def\BibTeX{{\rm B\kern-.05em{\sc i\kern-.025em b}\kern-.08em
    T\kern-.1667em\lower.7ex\hbox{E}\kern-.125emX}}
    
\usepackage[left=0.64in, right=0.64in, top=0.75in]{geometry}    
\newcommand\semiHuge{\@setfontsize\semiHuge{22.72}{27.38}}
\renewcommand{\paragraph}{\@startsection{paragraph}{4}{\z@}%
	{-2.00ex\@plus -1ex \@minus -.2ex}%
	{0.5ex \@plus .2ex}%
	{\normalfont\normalsize}}
\begin{document}

\title{{Optimal Time Allocation in VANETs Advertising: A Price-based Approach using 
Stacklberg Game}\thanks{This work was supported in part by the National Science Foundation under grants ECCS-1444009 and CNS-1824518.}  }
	
\author{\IEEEauthorblockN{Ali Rahmati, Seyyedali Hosseinalipour, and Huaiyu Dai}
\IEEEauthorblockA{{$^*$Department of Electrical and Computer Engineering}, 
{NC State University,}
Raleigh, NC, US \\
Email:\texttt{ \{arahmat, shossei3, hdai\}@ncsu.edu}}
}

% \title{Optimal Time Allocation in VANETs Advertising: A Price-based Approach using 
% Stacklberg Game}

% \author{\IEEEauthorblockN{Ali Rahmati}
% \IEEEauthorblockA{\textit{ECE Department} \\
% \textit{North Carolina State University}\\
% Raleigh, USA \\
% arahmat@ncsu.edu}
% \and
% \IEEEauthorblockN{Seyyedali Hosseinalipour}
% \IEEEauthorblockA{\textit{ECE Department} \\
% \textit{North Carolina State University}\\
% Raleigh, USA \\
% shossei3@ncsu.edu}
% \and
% \IEEEauthorblockN{Huaiyu Dai}
% \IEEEauthorblockA{\textit{ECE Department} \\
% \textit{North Carolina State University}\\
% Raleigh, USA \\
% hdai@ncsu.edu}
% }
\maketitle

%%%%%%%%%%%%%%%%%%%%%%%%%%%%%
% Abstract
%%%%%%%%%%%%%%%%%%%%%%%%%%%%%
\begin{abstract}
Vehicular ad-hoc networks (VANETs) have recently attracted a lot of attention due to their immense potentials and applications. Wide range of coverage and accessibility to end users make VANETs a good target for commercial companies. In this paper, we consider a scenario in which advertising companies aim to disseminate their advertisements in different areas of a city by utilizing VANETs infrastructure. These companies compete for renting the VANETs infrastructure to spread their advertisements. We partition the city map into different blocks, and consider a  manager for all the blocks who is in charge of splitting the time between interested advertising companies. Each advertising company (AdC) is charged proportional to the allocated time. In order to find the best time splitting between AdCs, we propose a Stackelberg game scheme in which the  block manager assigns the companies to the blocks and imposes the renting prices to different  companies in order to maximize its own profit. Based on this,  AdCs request  the amount of time  they desire to rent the infrastructure in order to maximize their utilities. To obtain the Stackelberg equilibrium of the game, a mixed integer non-linear optimization problem is solved using the proposed optimal and sub-optimal algorithms. The simulation results demonstrate that the sub-optimal algorithm  approaches the optimal one in performance with lower complexity.
\end{abstract}
\IEEEpeerreviewmaketitle

%%%%%%%%%%%%%%%%%%%%%%%%%%%%%
% Introduction
%%%%%%%%%%%%%%%%%%%%%%%%%%%%%
\vspace{-2mm}
\section{Introduction}
\noindent Recently, vehicular ad-hoc networks (VANETs) have attracted considerable interests due to their emerging applications such as collision avoidance, route finding, multi-hop transmission and traffic control \cite{ref:VANETsurvay, ferdowsi2018robust, rahmati2019dynamic,ferdowsi2017deep}. These networks are built upon two main components which are vehicles and infrastructure. The infrastructure in VANETs mainly consists of  roadside units (RSUs)  used for communication to vehicles. In addition, the vehicles are both capable of communicating with RSU and spreading the information among themselves with low latency~\cite{parvez2018survey}.

These networks are mainly developed for safety purposes such as collision avoidance. However, such networks with great capability of access to end users have always been enticing for  advertising companies (AdCs) \cite{8254662,8253975}. Consider a scenario in which AdCs rent VANETs RSUs for some period of time in order to spread their commercials among the vehicles (infrastructure-to-vehicle (I2V) mode).  This scenario can potentially lead to a substantial amount of profit for an  AdC. Moreover, the VANETs manager can rely on these networks as a potential source of profit by charging the AdCs accordingly. 

Until now, most of the literature in VANETs is dedicated to designing efficient routing protocols, describing mobility models, and improving routing efficiency using clustering based approaches\cite{ref:SurvayHuristicRouting,ref:SurveyClustering,ref:Surveyrouting,ref:SurveyMobility}. Nevertheless, there have also been some works on VANETs advertisement in which the most related ones are
~\cite{ref:1,ref:2,ref:3,8254662,8253975}.  In~\cite{8254662}, an auction-based framework is proposed to rent the blocks to the eager AdCs.  In~\cite{8253975}, an online-learning algorithm is proposed to perform block selection for AdCs with no prior information about the density of the vehicles inside the blocks. Although similar studies may exist in the literature, there is a lack of study focusing on modeling the interactions between AdCs and the manager which is in charge of controlling VANETs RSUs. This paper is prompted mainly due to this shortage. In~\cite{ref:1},  a scheme is proposed in which buses are the seeds of advertisement spreading in VANETs. It is assumed that all buses have the same set of advertising segments. In this scheme, buses have to find the most valuable segments for their surrounding vehicles and spread these segments. Afterwards, the authors use the coalition game in order to guide private vehicles to build broadcast coalitions when there is no bus available in the area. In \cite{ref:2}, a scenario is considered in which public transportation vehicles are the sources of advertisement spreading. The goal in that study is to find the most influential seeds to maximize the spreading range of the advertisement in the network. They have done an experimental study on two cities in China in order to obtain the temporal correlations for social centralities of vehicles to find the best initial seeds. In~\cite{ref:3}, a framework is proposed for \textit{virtual marketing} in which vehicles communicate  to find the possible matches between queries and demands.

One of the key challenges in the VANET advertisement scheme is managing the trading scheme between AdCs and the manager. The manager has to decide how to split the time between eager AdCs which are interested in the same area of a city for the purpose of advertising spreading using RSUs. More precisely, the manager has to assign the existing RSUs to the AdCs and split the time of utilization between them while making a satisfactory profit.

In this paper, we consider a scenario in which AdCs can rent the RSUs from the manager for limited time periods. The goal is to find the optimal leasing times for these AdCs given their utilities in such a way that the manager makes the maximum profit while fulfilling the companies' demands. Here, we propose a model based on the Stackelberg game which has a good match for the problem in hand and develop an analytic framework for the problem. In order to obtain the Stackelberg Equilibrium of the game, two algorithms are proposed. One of them is capable of solving the problem optimally with a high computation complexity. As an alternative, another algorithm is proposed which has a lower computation complexity with a sub-optimal solution. In simulation, good performance of the sub-optimal algorithm is revealed as compared to the optimal algorithm.

The rest of the paper is organized as follows. The system model and the VANET advertisement scheme are presented in Section~\ref{sec:2}. In addition,  the Stackelberg game basic definitions and adaptation to the VANET advertisement problem are presented as well. The details for obtaining the  Stackelberg equilibrium is discussed in Section~\ref{SE12}. Simulation results are depicted in Section~\ref{sec:4}. Finally, Section~\ref{sec:5} concludes the paper.

\section{System Model} \label{sec:2}
\subsection{Scenario Description}
\noindent 
The infrastructure in VANETs consists of different RSUs used  to both spread information to vehicles and gather information from them. In the mobile advertising scheme, a vehicle can receive the advertisements from one of these RSUs in the I2V mode. 
Consider a scenario in which different AdCs are willing to disseminate their advertisements through a city using the VANETs infrastructure. We grid the city map into different blocks where each block corresponds to an area of the city. In this scenario, an AdC might be interested in disseminating its advertisements through specific blocks. 
 Note that each block should be under the coverage of at least one of the RSUs so that the advertisements can be spread among the vehicles inside the block. We consider a block manager who is in charge of allocating the RSUs between AdCs. The block manager charges the AdCs upon using the infrastructure accordingly.

 Here, we partition the time into multiple non-overlapping time-batches each with a certain duration. At each time-batch, it is assumed that there are a finite number of AdCs willing to rent the RSUs in each block. At the beginning of each time-batch, the block manager splits the available time between the available AdCs and let each of them  spread its advertisement during its allocated portion. This splitting should be performed in such a manner that jointly maximizes the block manager and AdCs' utility. The competition begins when there are more than one AdC who are interested in a certain block in the same time-batch. In this paper, we aim to answer the following question: \textit{How much time should be allocated to each of these AdCs in order to maximize the block manager's profit, and what is the best strategy for AdCs so as to maximize their profits?}

\begin{figure}[!t]
\includegraphics[width=7.6cm,height=8cm]{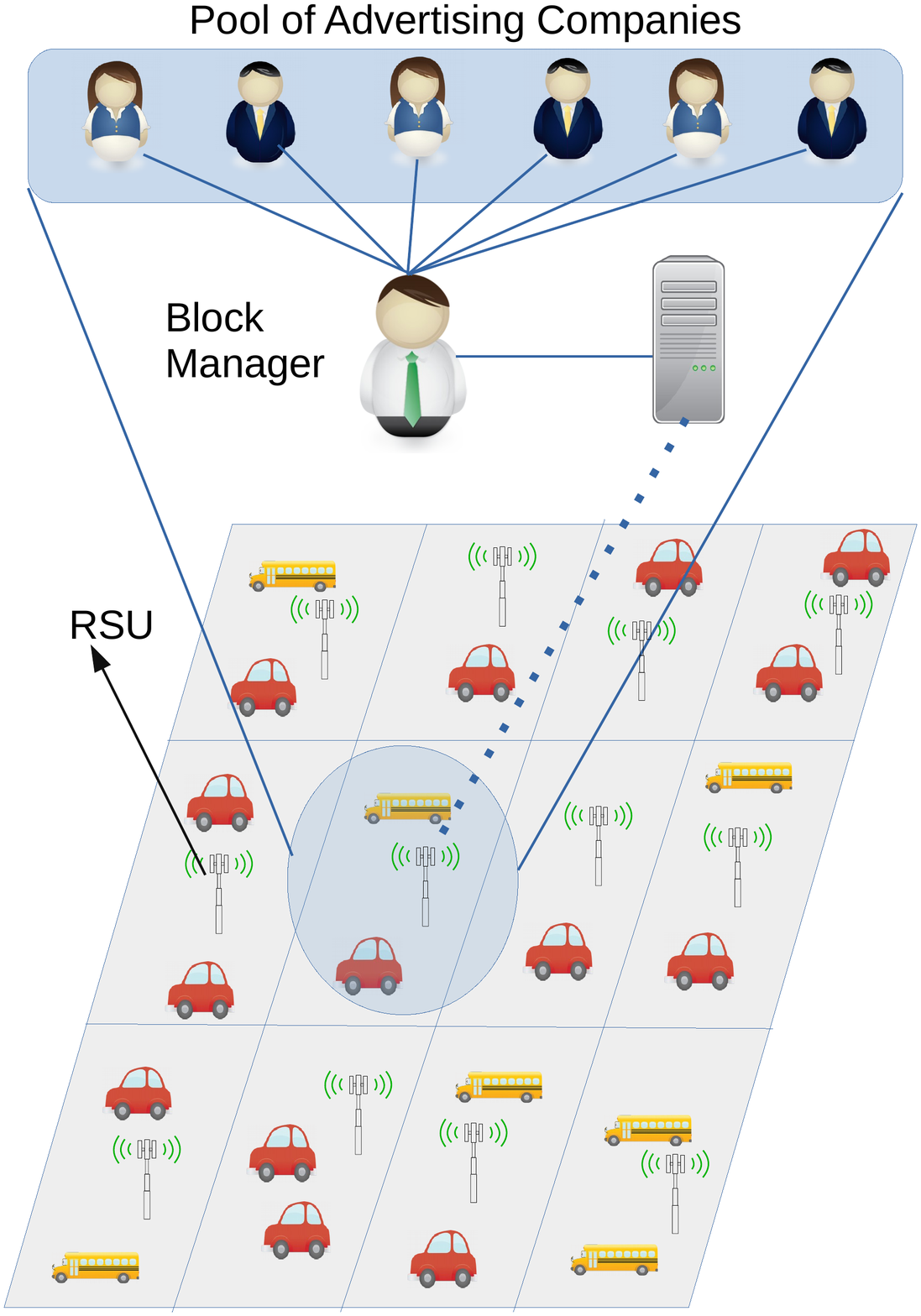}
\centering
\caption{A typical schematic for VANET.}
\label{fig:1}
\end{figure}

A schematic of our model is depicted in Fig.~\ref{fig:1}. The block manager assigns  the AdCs to some of the blocks and imposes the prices for each of the participants available in the pool of that block. This price demonstrates the cost per unit time which the corresponding AdC has to pay in order to rent the infrastructure. According to this price, each of the AdCs requests the amount of time they need in order to rent the infrastructure for advertisement. To model this  scenario, we exploit the Stackelberg competition scheme.

\subsection{Stackelberg Game and VANET Advertising Problem} \label{sec:3}
\noindent Consider a case in which the city is gridded into $M$ blocks, and there are $N$ AdCs willing to rent the blocks. Here, for the $j^{th}$ block, we partition the time into multiple non-overlapping time-batches each with the duration of $T_j$. In this manner, the $k^{th}$ time-batch captures the time interval of $t\in[(k-1)\; T_j, kT_j)$. Without loss of generality, we focus on the $k^{th}$ time-batch of all the blocks where $k$ is a fixed non-negative integer. We assume that the vehicles density is constant during each time-batch. Each AdC $i$ is interested in renting at most $M_i$  blocks for its advertisement. Through the rest, we refer to AdCs as followers and the block manager as the leader to be consistent with the framework of the Stackelberg game theory. 
We define the $i^{th}$ follower's satisfaction function in block $j$ as:
\begin{equation}\label{yek}
S(t_{i,j})=\lambda_{i,j} \left( 1-e^{-\alpha_{j}\frac{t_{i,j}}{T_{i,j}} } \right),~~ \forall i,j
\end{equation}
where $t_{i,j} $ is the requested time by  follower $i$ at block $j$, $T_{i,j}$ is a parameter subject to design,  $\alpha_{j} \ge 1$ is the vehicle density at block $j$, and $\lambda_{i,j}$ is the maximum satisfaction value\footnote{ $\lambda_{i,j}=0$ implies that follower $i$  is not interested in block $j$ for some reason.}. From~\eqref{yek}, it can be seen that the  satisfaction function is equal to zero if $t_{i,j}=0$, and it is an increasing function with respect to $t_{i,j}$. However, the rate of increase becomes smaller as $t_{i,j}$ gets larger (diminishing returns), and eventually saturates at the maximum value of $\lambda_{i,j}$. Here, we introduce the binary variable $a_{i,j}$ in order to specify whether the AdC $i$ advertises in block $j$ ($a_{i,j}=1$) or not ($a_{i,j}=0$).
 Moreover, the followers' utility function can be defined as the difference between their satisfaction and the charged price by the leader. The utility function of each AdC $i$ is given by:
\begin{equation}\label{panj1}
U_{i}( \mathbf{t}_i, \mathbf{p}_i, \mathbf{a}_i)=\sum_{j=1}^{M} a_{i,j}(S(t_{i,j})-p_{i,j} t_{i,j}), ~~\forall i,
\end{equation}
where $p_{i,j}$ is the unit price imposed to the follower $i$ for block $j$, $\mathbf{t}_i=[t_{i,1}, t_{i,2}, \dots , t_{i,M}]^T,  \mathbf{p}_i=[p_{i,1}, p_{i,2}, \dots , p_{i,M}]^T, \mathbf{a}_i=[a_{i,1}, a_{i,2}, \dots , a_{i,M}] $.
 Each follower aims to maximize its utility selfishly. Thus, given the prices and assignment variables by the leader, the  \textit{follower $i$'s sub-game}  can be written as:
\begin{align}\label{eq_20}
 &    \hspace{- 22mm} (\mathcal{P}_1): ~~~~~~~~  \max_{\mathbf{t}_i}
 & &  \hspace{- 14mm} U_{i}( \mathbf{t}_i, \mathbf{p}_i, \mathbf{a}_i), \nonumber \\
 & {\text{ s.t.}}
& & \hspace{-14mm} t_{i,j} \ge 0, ~~ \forall j. \nonumber
\end{align}
where $t_{i,j}$ is requested by the AdC, while the manager decides to accept the request or not in each block. 
 On the other hand, the leader attempts to maximize its utility which is the total revenue obtained from the followers. This is called the \textit{leader's sub-game} which can be written as:
\begin{align}
&\hspace{-12mm} (\mathcal{P}_2): ~~~~ \max_{\mathbf{p}, \mathbf{a}}
& &   \hspace{-4mm}  U^M(\mathbf{t}, \mathbf{p}, \mathbf{a}) =\sum_{i=1}^{N}\sum_{j=1}^{M} a_{i,j} p_{i,j} t_{i,j} \nonumber \\
& {\text{ s.t.}}
& & \hspace{-4 mm} C1: ~~ p_{i,j} \ge 0, ~\forall  i,j  \nonumber\\
& && \hspace{-4 mm} C2: ~~\sum_{i=1}^{N}  a_{i,j} t_{i,j} \le T_j, \forall j  \nonumber\\
&&& \hspace{-4 mm}C3: ~~ \sum_{j=1}^{M} a_{i,j} \le M_i, ~~\forall i  \nonumber \\
&&& \hspace{-4 mm}C4: ~~ a_{i,j} \in \{0,1\}, \nonumber 
\end{align}
where $\mathbf{t}, \mathbf{p}, \mathbf{a}$ are the matrices with elements $t_{i,j},$ $ p_{i,j}$, $ a_{i,j},~\forall i,j$, respectively. As can be seen, the leader aims to maximize its revenue by assigning the AdCs to blocks and determining the prices, while the constraint on the total available time is  met.  

\section{The Stackelberg Equilibrium (SE)}\label{SE12}
\noindent In this section, we aim to find the SE of the proposed game. 
\begin{definition}
Let $\mathbf{r}$ denote the strategy set spanned by $\mathbf{p}$ and~$\mathbf{a}$. Moreover, let $\mathbf{r}^*$ and $\mathbf{t}^*$ denote the   optimal strategy for the leader, and followers, respectively. Then, the point $(\mathbf{r}^*,\bm{t}^*_i)$  is a SE if the following conditions are satisfied \cite{rahmati2017price}:
\begin{equation}\label{123}
\hspace{-16mm} U (\mathbf{r}^*,\mathbf{t}^*)\ge U (\mathbf{r},\mathbf{t}^*), ~\forall ~\mathbf{r},
\end{equation}
\begin{equation}\label{124}
U_{i} (\mathbf{t}^*,\mathbf{r}^*)\ge U_{i} (\mathbf{t}_{i},\mathbf{t_{-i}}^*,\mathbf{r}^*),\ \ \forall~ \mathbf{t}_{i}\geq 0.
\end{equation}
\noindent where $\mathbf{t_{-i}}^*$ is the optimal time matrix including the optimal time vector for all the followers except follower $i$ \cite{fudenberg1991game}. 
\end{definition}
\noindent In what follows, we aim to obtain the SE of the proposed game using backward induction \cite{rahmati2015price,rahmati2016price}.

\subsection{Followers' sub-game}
\begin{lemma}
Given the allocated vector $\mathbf{p}_i$ and block assignment vector $\mathbf{a}_i$ for the $i^{th}$ follower, its sub-game has a global optimal solution given by:
\begin{equation}\label{panj3}
t_{i,j}^*(p_{i,j})=  \left( \frac{T_{i,j}}{\alpha_{j}} \ln{\frac{\lambda_{i,j} \alpha_{j}}{T_{i,j} p_{i,j}}} \right)^+, \ \ \forall j,
\end{equation}
where $(x)^+ \triangleq \max(x,0)$.
\end{lemma}
\begin{IEEEproof}
Given $\mathbf{a}_i$ and $\mathbf{p}_i$, we need to obtain  the Hessian matrix of $U_{i}( \mathbf{t}_i)$ for each follower $i$ which is only a function of $\mathbf{t}_i$.
For all $n\ne m$, we have: 
\begin{equation}\label{125}
\frac{\partial^2 U_{i}(\mathbf{t}_i) }{\partial t_{i,n} t_{i,m}}=0,~~ \forall i.
\end{equation}
The diagonal elements of the Hessian matrix are: 
\begin{equation}\label{126}
\frac{\partial^2 U_{i}({\mathbf{t}_{i}}) }{\partial t_{i,j}^2 }=-\lambda_{i,j} \frac{\alpha^2_{i,j}}{T_{i,j}^2}e^{-\alpha_i\frac{t_{i,j}}{T_{i,j}}}, \ \ \forall {j}.
\end{equation}
Thus, the Hessian matrix is negative semi-definite and the objective function is concave. As a result, the followers' sub-game is a convex optimization problem. The best strategy for each follower $i$ can be obtained by setting the first order derivative of the utility function with respect to $t_{i,j}$ equal to zero.
After some manipulations, the optimal solution for $t_{i,j}$ can be written as in \eqref{panj3}.
\end{IEEEproof}

Setting the optimal solution for each $t_{i,j}, ~ \forall i,j$ greater than or equal to zero, the following condition can be obtained:
\begin{equation}\label{hasht}
 p_{i,j} \le {\frac{\lambda_{i,j} \alpha_{j} }{T_{i,j} }} , ~~\forall i.
\end{equation}
If this constraint is not met for a follower,  its corresponding term in utility function in \eqref{panj1}  takes a negative value. In this situation,  follower $i$ prefers to back off and resign the game.

\subsection{Leader's sub-game}
\noindent Substituting the optimal solution of the followers' sub-game into  the leader's sub-game and imposing the constraints in \eqref{hasht}, one can obtain the following optimization problem $(\mathcal{P}_3)$ for the leader:
\begin{align}
&   \max_{\mathbf{p}, \mathbf{a}}
& &   \hspace{0mm} U(\mathbf{p}, \mathbf{a}) =\sum_{i=1}^{N}\sum_{j=1}^{M} a_{i,j} \left(\frac{p_{i,j} T_{i,j}}{\alpha_{j}} \ln{\frac{\lambda_{i,j} \alpha_{j}}{T_{i,j} p_{i,j}}} \right) \nonumber \\
& {\text{ s.t.}}
& & \hspace{0 mm} C1, C3, C4, \nonumber\\
& && \hspace{0 mm}\hat C2:~~\sum_{i=1}^{N}  a_{i,j} \left( \frac{T_{i,j}}{\alpha_{j}} \ln{\frac{\lambda_{i,j} \alpha_{j}}{T_{i,j} p_{i,j}}} \right) \le T_j, \forall j  \nonumber\\
&&&  C5:~~ p_{i,j} \le {\frac{\lambda_{i,j} \alpha_{j} }{T_{i,j} }} , ~~\forall i,j.  \nonumber
\end{align}
As  can be seen, the optimization problem $(\mathcal{P}_3)$ is a non-convex mixed integer nonlinear optimization problem (MINLP), and the objective function of the problem is non-convex. The non-convexity arises due to the multiplication of  $a_{i,j}$ by $p_{i,j}$.
Thus, this problem can not be solved using a tractable method and needs reformulation.
In the following section, we propose two algorithms capable of solving the leader optimization problem.
\section{Solution for the Leader Sub-game}

 \subsection{Optimal solution using Generalized Bender's Decomposition (GBD) Algorithm}
 \noindent  The GBD algorithm consists of solving an alternating sequence of relaxed problems including mixed integer linear problems
(MILP) called master problems and convex optimization problems called primal problems. At each iteration, the algorithm generates an upper bound and a lower bound on the
MINLP solution. The lower bound of the optimal solution LB$^{(l)}$ can be obtained from the  objective value of the  primal problem, while the upper bound of the optimal solution UB$^{(l)}$ is the objective value of the master problem. The master problem is an integer linear problem that can be solved efficiently using standard optimization toolboxes, e.g, MOSEK. As the iterations proceed,
two sequences of updated upper bounds and lower bounds converge to the optimal solution in a finite
number of iterations \cite{floudas1995nonlinear}. More details are given below.

\subsubsection{Problem Reformulation}
The optimization problem $(\mathcal{P}_3)$ is non-convex due to the multiplicative term of $a_{i,j}p_{i,j}$.
To apply the GBD algorithm, we recast  $(\mathcal{P}_3)$ to an equivalent
 convex MINLP optimization problem  as $(\mathcal{P}_4)$:
\begin{align}
&   \max_{{\bf{p}},{\bf{a}}}
& &   \hspace{0mm} U({\bf{p}},{\bf{a}}) =\sum_{i=1}^{N}\sum_{j=1}^{M}  \frac{p_{i,j} T_{i,j}}{\alpha_{j}} \ln{\frac{\lambda_{i,j} \alpha_{j}}{T_{i,j} p_{i,j}}} \nonumber  \\
& {\text{ s.t.}}
& & \hspace{0 mm} C1, C3, C4, C5  \nonumber\\
& && \hspace{0 mm} \tilde C2: ~~\sum_{i=1}^{N}   \frac{T_{i,j}}{\alpha_{j}} \ln{\frac{\lambda_{i,j} \alpha_{j}}{T_{i,j} p_{i,j}}} \le T_j, \forall j  \nonumber\\
& && \hspace{0 mm} C6: ~~\frac{T_{i,j}}{\alpha_{j}} \ln{\frac{\lambda_{i,j} \alpha_{j}}{T_{i,j} p_{i,j}}}   \le a_{i,j} T_j, \forall j.  \nonumber 
\end{align}
As can be seen in $(\mathcal{P}_3)$, each binary variable $a_{i,j}$ is multiplied by a term in the objective function and  $\hat C2$. The role of this binary variable is to make such terms zero when $a_{i,j}=0$ and leave them as they are when $a_{i,j}=1$. In $(\mathcal{P}_4)$, it can be seen that variables $a_{i,j}$  are not multiplied by the corresponding terms in the  objective function and  $\tilde C2$. Instead of this, constraint $C6$ is added. 
In constraint $C6$, if $a_{i,j}=0$, after some manipulations,  it can be simplified as ${{\lambda_{i,j} \alpha_{j} }/{T_{i,j} }} \le p_{i,j}$. Together with $C5$, the only possible value for $p_{i,j}$ is $ p_{i,j}={{\lambda_{i,j} \alpha_{j} }/{T_{i,j} }}$, which forces the  terms associated with follower $i$ and block $j$ in the objective function and $\tilde C2$  to become  zero. Otherwise, if $a_{i,j}=1$,  $C6$ for the follower $i$ in block $j$ becomes redundant since $T_j$ is an upper bound for it based on $\hat C2$. We can now observe that the optimization problem  $(\mathcal{P}_4)$ is a convex MINLP. In addition,  the continuous variables ${\bf{p}}$ and the binary variables ${\bf{a}}$ are decoupled. Hence, we can deploy GBD to solve~$(\mathcal{P}_4)$.
 \begin{algorithm}[t]
	
	\DontPrintSemicolon
	Initialize $\mathbf{a}^{(0)}$,
	
	Set convergence error $\epsilon$, ${\mathrm{Upper~ bound~UB}}^{(l)}=\infty, \ \mathrm{Lower~ bound~LB}^{(l)}=0 $.
	
	$ l \gets  1$
	
	\While{ $|{\mathrm{UB}}^{(l)}-\mathrm{LB}^{(l)}| \ge \epsilon$}{
		
		Solve primal problem and obtain $ {\bf {p}}^*,  \bm\beta^{(l)*}, \bm \gamma^{(l)*}, \bm \nu^{(l)*}$ and the lower bound, LB$^{(l)}$.
		
		Solve the master problem and obtain $\delta^* $, $ \mathbf{a}^{(l)*}$, and the upper bound UB$^{(l)}$.
		
		$ l \gets  l+1$ }

	\caption{{Generalized Bender's Decomposition}}
	\label{alg}
\end{algorithm}
\subsubsection{Master and Primal Problems}
Here, we decompose $(\mathcal{P}_4)$ into two sub-problems.
The Primal and Master problems are presented as follows.

\text{\bf{Primal Problem ($l$-th iteration):}} For the given optimal binary variables at iteration $l-1$, ${\bf{a}}^{(l-1)*}$, the primal problem can be formulated as follows:
\begin{align}
&   \max_{\bf{p}}
& &   \hspace{0mm} U({\bf{p}}) =\sum_{i=1}^{N}\sum_{j=1}^{M}  \frac{p_{i,j} T_{i,j}}{\alpha_{j}} \ln{\frac{\lambda_{i,j} \alpha_{j}}{T_{i,j} p_{i,j}}}  \nonumber \\
& {\text{ s.t.}}
& & \hspace{0 mm} C1, \tilde C2, C5,  \nonumber\\
& && \tilde C6:~~\hspace{0 mm} \frac{ T_{i,j}}{\alpha_{j}} \ln{\frac{\lambda_{i,j} \alpha_{j}}{T_{i,j} p_{i,j}}}   \le a_{i,j}^{(l-1)*} T_j, ~\forall i,j.  \nonumber
\end{align}
This problem can be solved using standard convex optimization methods including interior point algorithm \cite{haykin2005cognitive}. 

 \text{\bf{Master Problem ($l$-th iteration):}} The master problem, which provides the upper bound of the solution, is formulated based on the Lagrangian of the primal problem. The
Lagrangian of the primal problem can be written as:
\begin{align}\nonumber
 &   
 & &\hspace{0mm}   \mathcal{L}({\bf {p}}, {\bf {a}},   \bm \beta, \bm \gamma, \bm \nu)=  
\sum_{i=1}^{N}\sum_{j=1}^{M}  \frac{p_{i,j} T_{i,j}}{\alpha_{j}} \ln{\frac{\lambda_{i,j} \alpha_{j}}{T_{i,j} p_{i,j}}} \nonumber\\
& & & \hspace{-4mm}  - \sum_{j=1}^{M} \beta_j \left( \sum_{i=1}^{N}   \frac{ T_{i,j}}{\alpha_{j}} \ln{\frac{\lambda_{i,j} \alpha_{j}}{T_{i,j} p_{i,j}}}  - T_j \right)   \nonumber \\
& & &\hspace{-4mm}  -\sum_{n=1}^{N} \sum_{j=1}^{M} \gamma_{i,j}\left (p_{i,j} - {\frac{\lambda_{i,j} \alpha_{j} }{T_{i,j} }} \right ) \nonumber  \\
& & &\hspace{-4mm}  -\sum_{n=1}^{N} \sum_{j=1}^{M} \nu_{i,j}\left ( \frac{ T_{i,j}}{\alpha_{j}} \ln{\frac{\lambda_{i,j} \alpha_{j}}{T_{i,j} p_{i,j}}}  - a_{i,j} T_j \right ), \nonumber
\end{align}
where $ \bm \beta, \bm \gamma, \bm \nu$ represent the Lagrange multipliers associated with constraints $\tilde C2, C5, \tilde C6$, respectively.
Assume the optimal values of the Lagrange multipliers of the primal problem are  $ \bm \beta^{(l)*}, \bm \gamma^{(l)*}, \bm \nu^{(l)*}$ at iteration $l$. Hence, the master problem can be formulated as:
\begin{align}\label{eq5}
 &  
 & &  \hspace{-1cm}\max_{\delta, {\bf{a}}}\hspace{0.3cm} \delta \\
 & \hspace{-1cm} {\text{ s.t.}}
&  &  \hspace{-2cm} \delta \le \mathcal{L}(  {\bf {p}}^*, {\bf {a}}, \bm \beta^{(l)*}, \bm \gamma^{(l)*}, \bm \nu^{(l)*}), ~~\forall l   \nonumber \\
& & &  \hspace{-2cm} C3, C4. ~~\nonumber
\end{align}
Given the optimal values of ${\bf{p}}^*$ and optimal values of Lagrange multipliers, the master problem is a MILP which  can be solved optimally by any standard optimization toolboxes, e.g., MOSEK~\cite{mosek2010mosek}. The GBD algorithm for this problem is described in {\bf{Algorithm~1}}.

\subsection{Sub-optimal Heuristic Solution}
Since by increasing the number of users $N$ and the number of blocks $M$ the computation complexity of the GBD algorithm increases considerably, we need to come up with a more practical solution with a lower complexity. The basic idea for our proposed heuristic algorithm is to solve  $(\mathcal{P}_3)$ for a given assignment variable ${\bf{a}}$ as a convex optimization problem with respect to ${\bf{p}}$  in one shot. To obtain this assignment variable, we first consider an extreme case where $T_j \to \infty, ~\forall j$, for which constraint $\hat C2$ is redundant. In this case, the objective function in  problem $(\mathcal{P}_3)$ is decoupled in terms of $\{p_{i,j}\}$. As a result, the corresponding optimal value for $ p_{i,j},~ \forall i, j$ can be directly obtained as:
\begin{equation}\label{dah}
\tilde p_{i,j}=\frac{\alpha_{j}\lambda_{i,j}}{T_{i,j} e}, ~ \forall i, j.
\end{equation}
Substituting \eqref{dah} into $(\mathcal{P}_3)$, problem  $(\mathcal{P}_3)$ can be reduced as:
\begin{align}
&\hspace{-12mm} (\mathcal{P}_5): ~~~~  \max_{\mathbf{a}}
& &   \hspace{-1cm}\sum_{i=1}^{N}\sum_{j=1}^{M} a_{i,j} \lambda_{i,j} \nonumber \\
& {\text{ s.t.}}
& & \hspace{-1cm }  C3, C4. \nonumber
\end{align}
After obtaining the optimal assignment from $(\mathcal{P}_5)$ as ${\bf{a}}^*$, by substituting it into $(\mathcal{P}_3)$,  the  optimization problem $(\mathcal{P}_6)$ over $\mathbf{p}$ can be obtained:
\begin{align}
&   \max_{\mathbf{p}}
& &   \hspace{0mm} U(\mathbf{p}, \mathbf{a}^*) =\sum_{i=1}^{N}\sum_{j=1}^{M} a_{i,j}^* \left(\frac{p_{i,j} T_{i,j}}{\alpha_{j}} \ln{\frac{\lambda_{i,j} \alpha_{j}}{T_{i,j} p_{i,j}}} \right) \nonumber \\
& {\text{ s.t.}}
& & \hspace{0 mm} C1, C3, C4, C5 \nonumber\\
& && \hspace{0 mm}\tilde C2:~~\sum_{i=1}^{N}  a_{i,j}^* \left( \frac{T_{i,j}}{\alpha_{j}} \ln{\frac{\lambda_{i,j} \alpha_{j}}{T_{i,j} p_{i,j}}} \right) \le T_j, \forall j  \nonumber
\end{align}
Since the above optimization problem is convex, it can be solved using standard convex optimization toolboxes such as CVX.
The solution of the above optimization problem provides a sub-optimal solution with much lower computation complexity. It should be noted that for the GBD algorithm, at each iteration, one convex optimization and one MILP should be solved. Thus, assuming that it converges in $L$ iterations, we need to solve $L$ convex optimization problems and $L$ MILPs. In contrast, for our proposed sub-optimal algorithm, solving one integer linear programming $(\mathcal{P}_5)$ and one convex optimization problem $(\mathcal{P}_6)$ is sufficient. This leads to a significant decrease in the computational complexity.

\section{Simulation Results} \label{sec:4}
\noindent In simulations, we assume $N=5$ as the number of the AdCs and  $M=15$ as the number of blocks. The design parameter  for each AdC at each block $  T_{i,j},~ \forall i, j$ and the maximum satisfaction value $\lambda_{i,j}, ~\forall i, j$ are generated randomly based on the uniform distribution in the ranges $ [0,4], [0,10]$, respectively.  We consider a city map  divided
into 12 blocks (3 $\times$ 4 grid). In this map, the number of
vehicles in each block are  in the range of [1, 30]. The parameter $\alpha_{j}$ is proportional to the density of vehicles in each block.   As a baseline, we  compare our proposed schemes with random block assignment. Note that in the random block assignment scheme,   the blocks are assigned randomly in $(\mathcal{P}_3)$ and the problem $(\mathcal{P}_3)$ is solved  with respect to the price variables. It is assumed that $T_j=T,~\forall j$.

\begin{figure}
\centering
\includegraphics[width=10cm,height=6cm,keepaspectratio]{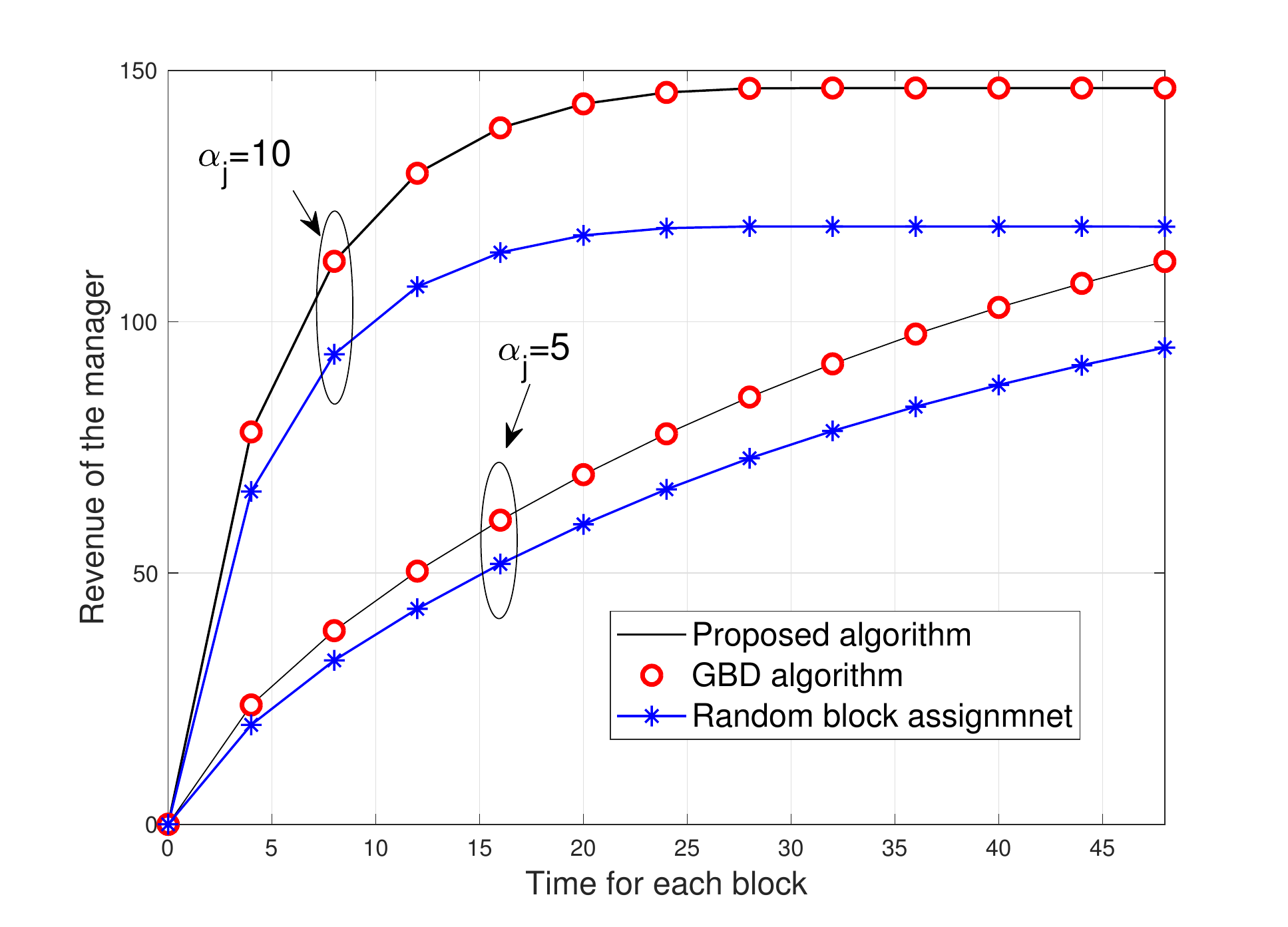}
\centering
\caption{Revenue of the manager vs. time for each block.}
\label{figyek}
\end{figure}

\begin{figure}
\centering
\includegraphics[width=10cm,height=6cm,keepaspectratio]{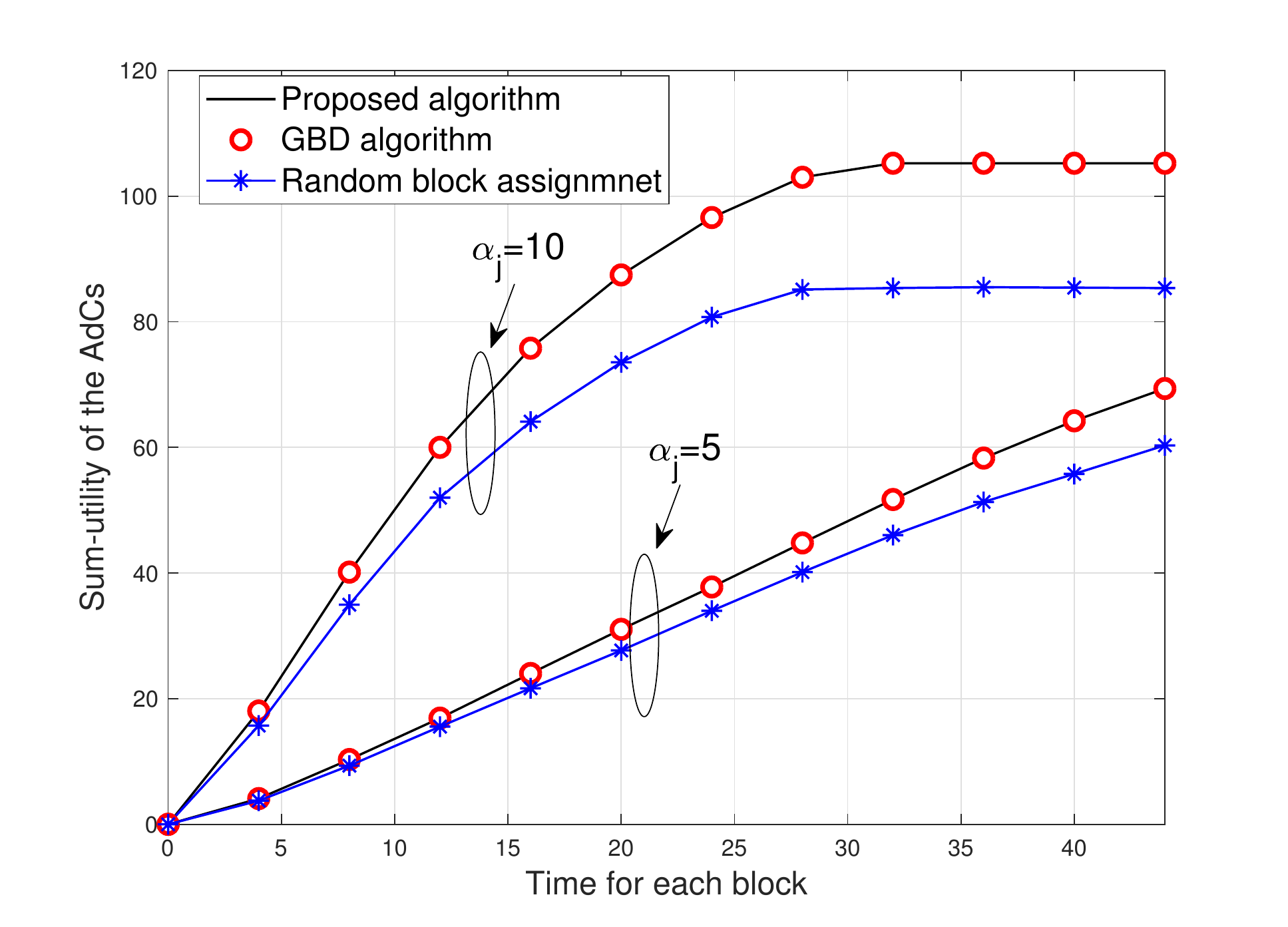}
\centering
\caption{Sum-utility of the AdCs vs. time for each block.}
\label{figdo}
\end{figure}

In Fig. \ref{figyek} and Fig. \ref{figdo}, the revenue of the block manager and sum-utility of the AdCs are shown for different  vehicle densities, respectively. The performance of the  GBD algorithm, the sub-optimal  algorithm and random block assignment with optimal  prices are depicted versus time. It can be seen that  GBD and the sub-optimal algorithm perform very closely to each other and their performance is much better than that of random block assignment. However, it should be noted that the computation complexity of  GBD is much higher than that of our sub-optimal algorithm. As a result, for large networks, it is better to use the sub-optimal algorithm enjoying lower complexity, which leads to almost the same revenue. It is also observed that by increasing the value of $T$ the revenue stops increasing. The reason is that the objective function in $\mathcal{P}_4$ is concave and has a global optimal solution. Thus, after a certain point, increasing  $T$ can not affect the solution of the problem anymore. That means the AdCs have essentially reached their maximum satisfaction and increasing the allocated time, won't further improve their utilities. Moreover, if the  vehicle density increases,   both the revenue of the  manager and the sum-utility of the AdCs  increase.

\begin{figure}
\centering
\includegraphics[width=10cm,height=6cm,keepaspectratio]{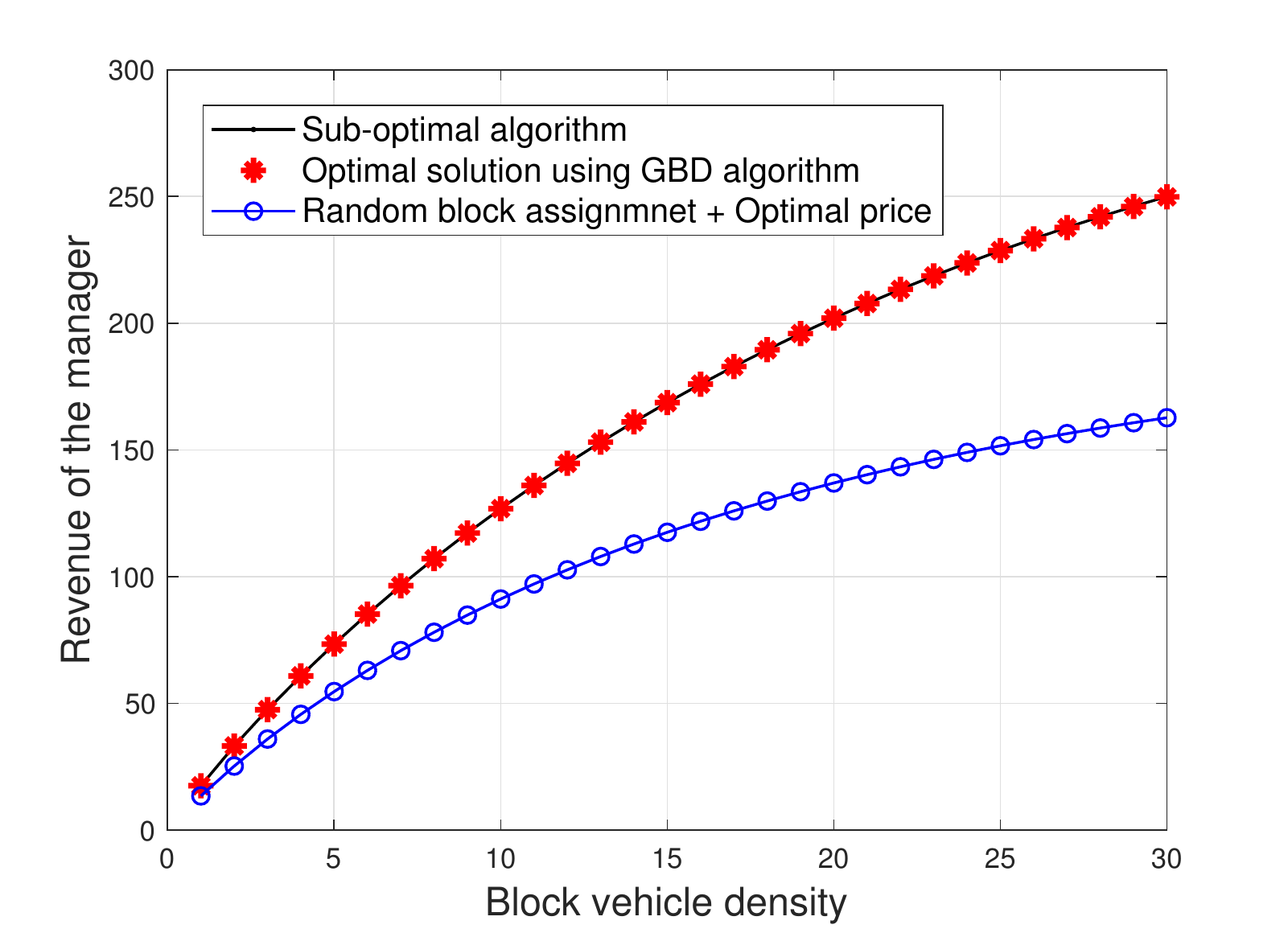}
\centering
\caption{Revenue of the manager vs. density of the vehicles.}
\label{figse}
\end{figure}

 In Fig.~\ref{figse}, the revenue of the manager is  depicted versus the density of the vehicles in each block.  Similar to the previous results, the GBD and the sub-optimal algorithms  exhibit almost identical performance, while the random assignment has much worse performance. As can be seen from Fig.~\ref{figse}, by increasing the density of the vehicles, the revenue of the manager increases due to the larger number of vehicles targeted for advertisement. Moreover, by increasing the density of vehicles, the gap between the random assignment algorithm and the other two becomes more obvious.

% \begin{figure*}
% \begin{multicols}{2}
%     	\centering

%     \includegraphics[width=7cm,height=6cm,keepaspectratio]{convs.eps}\par 
%     	\centering

%     \includegraphics[width=7cm,height=6cm,keepaspectratio]{convs.eps}\par
%     	\centering

%     \end{multicols}
% \begin{multicols}{2}
%     	\centering

%     \includegraphics[width=7cm,height=6cm,keepaspectratio]{convs.eps}\par
%     	\centering

%     \includegraphics[width=7cm,height=6cm,keepaspectratio]{convs.eps}\par
%     	\centering

% \end{multicols}

% \caption{caption here}
% \end{figure*}

% \begin{figure}[!t]
% 	\includegraphics[width=0.8\textwidth]{convs.eps}
% 	\centering
% 	\caption{Convergence of the GBD algorithm.}
% 	\label{figyek}
% \end{figure}
% \begin{figure}[!t]
% 	\includegraphics[width=0.8\textwidth]{manrev.eps}
% 	\centering
% 	\caption{Revenue of the manager vs. the time for each block.}
% 	\label{figdo}
% \end{figure}

\section{Conclusion} \label{sec:5}
\noindent In this paper, we studied the time allocation among AdCs in the context of VANET advertising. A block manager and multiple   AdCs are considered, where the manager aims to split the utilization time between AdCs so as to increase its revenue. We formulated the problem as a Stackelberg game aiming to jointly maximize the utility of the manager and AdCs. It is shown that the problem can be formulated as a mixed integer non-linear optimization problem. The SE of the proposed game is obtained using Generalized Bender’s Decomposition  after a reformulation. Prompted by the high complexity of the GBD algorithm, we proposed a sub-optimal algorithm for the problem, which enjoys lower complexity, and its performance is close to that of the GBD.

\bibliographystyle{ieeetr}
\bibliography{VANETstack}

\end{document}